# Thickening of liquids using copolymer grafted nanoparticles

Prama Adhya[1], Sachin M. B. Gautham[2,3], Tarak K Patra[2,3*], Manish Kaushal[1*], Titash Mondal[4*]

[1]Department of Chemical Engineering, Indian Institute of Technology Kharagpur, Kharagpur-721302, India
[2]Department of Chemical Engineering, Indian Institute of Technology Madras, Chennai-600036, India
[3]Center for Atomistic Modeling and Materials Design, Indian Institute of Technology Madras, Chennai-600036, India
[4]Rubber Technology Center, Indian Institute of Technology Kharagpur, Kharagpur-721302, India

**ABSTRACT**

Precise control over the flow behavior of liquids is a critical problem and is demanding for multifaceted applications. Introducing surface-engineered nanoparticles into the liquid can tune the flow behavior. However, the extent of tunability depends on the compatibility of the surface groups on nanoparticles and the liquid matrix. Herein, we report a strategy to synthesize solvophilic surface-engineered nanoparticles demonstrating its capability to control the flow property of chemically distinct liquids like silicone oil and polyethylene glycol. Silylated polyether amine-modified silica nanoparticles with polymer brush-like structure were selected as the model synthesized material. By combining rheological experiments with coarse-grained molecular dynamic (CGMD) simulations, we demonstrate that adding these grafted nanoparticles significantly impacts the viscosity of both liquids. The impact of grafting density on the rheological properties was also corroborated. Simulations indicate the formation of a diverse array of nanoparticle microstructures, including contact aggregates, bridged aggregates, and one-dimensional string-like structures that dictate the viscosity of the material. Furthermore, we show that copolymers with predominantly matrix-compatible moieties facilitate the intercalation of matrix molecules within the polymer brushes, enhancing interactions at the molecular level. This work provides fundamental insights into the design of composite materials with tunable rheological properties, offering new pathways for optimizing liquid formulations.

**Keywords:** Copolymer grafted nanoparticles, Rheological modifier, Molecular dynamics simulations, Nanoparticles aggregation, and Sequence control

Polymer-grafted nanoparticles (GNPs) are extensively studied for various applications, including personal care products, automotive technologies, healthcare, and energy storage devices.[1–5] They offer significant tunability in the properties of liquid systems. The primary objective of polymer grafting on the nanoparticles (NP) surface is to control the distribution of NPs within liquid matrices, thereby regulating the properties of the resulting composites.[1,2,6,7] Also, these GNPs exhibit rich phase behavior and a broad spectrum of tunable properties in their pristine states.[8–11] When incorporated into a polymeric liquid, their distribution is influenced by several key factors, including grafting density, the ratio of grafted chain length to free polymer chain length, and particle loading. The grafted polymers play a crucial role in modulating the effective NP-NP interaction within the matrix, ultimately dictating their spatial arrangement and its rheology. [12–15] It is worth mentioning that these studies have primarily used polymer grafts that are chemically similar to surrounding polymer matrices. However, for enhanced rheological performances and greater tunability, polymer grafts should ideally exhibit a broad range of compatibility with a diverse range of matrices, including chemically dissimilar ones. Achieving this level of versatility requires designing polymer grafts with complex physicochemical properties, such as block or statistical copolymers containing segments of distinct chemical characteristics. Prior experimental studies have explored diblock copolymers for improving NP dispersion, wherein one block is chemically similar to the free polymer matrix.[16,17] However, such diblock polymer grafts may not demonstrate solvophilicity towards chemically distinct liquids or effective thickeners for liquids. We hypothesize that achieving optimal control over NPs distribution and rheology necessitates monomer-level sequence variation in the graft chains. Previous CGMD simulations suggest that physical and chemical heterogeneities within the grafts can lead to a wide range of effective NP-NP interactions and spatial distributions.[18] Building on these insights, this study investigates how the sequence of copolymer grafts influences nanoparticle distribution within chemically distinct solvents like polyethylene glycol (PEG), silicone oil and thereby demonstrates its capability to provide superior control of rheology across a broad spectrum of liquid matrices.

Accordingly, we prepared silica-based GNPs with organic grafts, silylated polyether amine having two chemically dissimilar moieties viz., ethylene oxide (EO) and propylene oxide (PO) in a random sequence. Silica nanoparticles are of size 10 nm. We add these silica-based GNPs in two different liquids - PEG and silicone oil. Increase in the grafting density of the polymer chains over the silica NP monotonously increased the viscosity of both composites. A plateauing effect of the viscosity profile was noted when the silica NP of grafting density c.a. 2 chains/nm$^2$ was used. We perform CGMD simulations and systematically explore a few limiting cases alongside the random polymer grafting scenario to gain deeper insight into the underlying mechanisms. Unlike homopolymer grafts, which are chemically similar to the base matrix and typically ensure good dispersion of NPs at low grafting densities,[15,19,20] we find that copolymer grafting results in a diverse range of NP aggregates. Also, unlike homopolymer grafts, which often reduce the viscosity of the composites due to autophobic transition for the high grafting limit,[21] the current copolymer grafts are found to increase the viscosity irrespective of the base matrix for a wide range of grafting densities. The CGMD simulations strongly suggest that this viscosity increment originates from a wide range of NP aggregates, including contact and bridged GNPs aggregates. Our experiments show that the viscosity of a liquid can be enhanced by up to six orders of magnitude, depending on the grafting density and volume fraction of GNP fillers. The simulations reveal the mechanisms of liquid thickening in the presence of random copolymer GNPs, as seen in our experiments, and highlight the potential for extensive tunability of the rheological property through precise control of the monomer sequence in grafted chains.

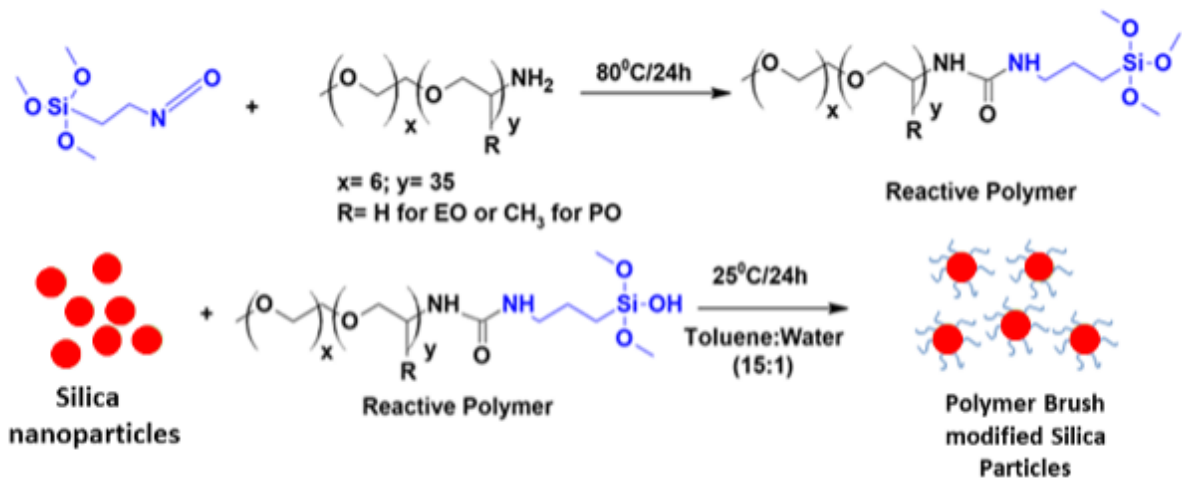

*Figure 1: Schematic representation of the pathway to synthesize the trimethoxy silane terminated polyethers followed by tethering of the functional copolymer over the silica particles at different grafting levels.*

We leverage a two-step process for surface modification of silica particles (**Supporting Information (SI), Section 2**). The reaction's first step involves synthesizing trimethoxy silane terminated polyethers. A monofunctional polyetheramine having a molar ratio of PO to EO as 10:31 is made to react with the 3-Isocyanatopropyltrimethoxysilane under optimized conditions. The 3-Isocyanatopropyltrimethoxysilane is preferentially made to react with the monofunctional polyetheramine over the silica NP to avoid multiple side reactions possible with surface active silanol group of the silica NP via the isocyanate group or the alkoxy group of the 3-Isocyanatopropyltrimethoxysilane. The reactive silane terminated polyether demonstrates a weight average molecular weight, $M_w \sim$ 2116 g/mol, number average molecular weight, $M_n \sim$ 2072 g/mol, and polydispersity index, PDI ($M_w/M_n$) $\sim$ 1.02 and is "*grafted to*" the silica particles following the procedure shown schematically in **Figure 1**. The ratio (w/w) of the surface modifier to the silica NPs is varied in the reaction course. The reactive polymer and the surface-modified silica NPs are thoroughly characterized using Fourier transform infrared spectroscopy (FTIR), thermogravimetric analysis (TGA) (**SI, Section 3, Figures S1a,b - S2a**). From the TGA data, the grafting density ($\Sigma$) is estimated using well-established protocols.[22,23] A broad spectrum of grafting density is obtained, ranging from $\sim$ 0.57 chain/nm$^2$ to 2.04 chains/nm$^2$ (**SI, Table S1**). An increment in the functional polymer concentration in the reaction media increases the grafting density. We can achieve a significantly higher grafting density than the values reported in the literature.[24–26] The static light scattering experiment is used to determine the radius of gyration ($R_g$) for the unperturbed functional polymer chains with an $R_g \sim$ 13.3 Å (**SI, Section 3, Figure S3a**). The reduced tether density ($\varphi$) is estimated from the various $\Sigma$ values and the $R_g$ using the equation $\varphi = \sum \pi R_g^2$. By comparing the $\varphi$ value reported in the literature,[27] a steady transition from "mushroom to brush regime" to a "true polymer brush regime" for the functionalized silica particles is noted with an increment in the concentration of the reactive polymer (**SI, Figure S3b**). It is also worth mentioning that, tethering of the polymer chains onto the NPs perturbs the polymer chain packing, and thereby impacts their thermal behavior. In this window of study, all the polymer brush modified silica NPs demonstrated an enhancement in the glass transition temperature ($T_g$) compared to the neat polymer (**SI, Figures S2b)**. Surface tethering warranted the loss of mobility of the end grafted polymer chains and also due to confinement, $T_g$ is noted to be increased. The $T_g$ of the tethered polymer chain is found to monotonically increase with an increment in the grafting density. Higher grafting density results in steric crowding, thereby exacerbated the entropy of the polymer chains, leading to an increase in the $T_g$ of the tethered chains.[28] The sample with highest grafting density registered a $T_g$ increased by ~ 7 K compared to the bulk polymer.

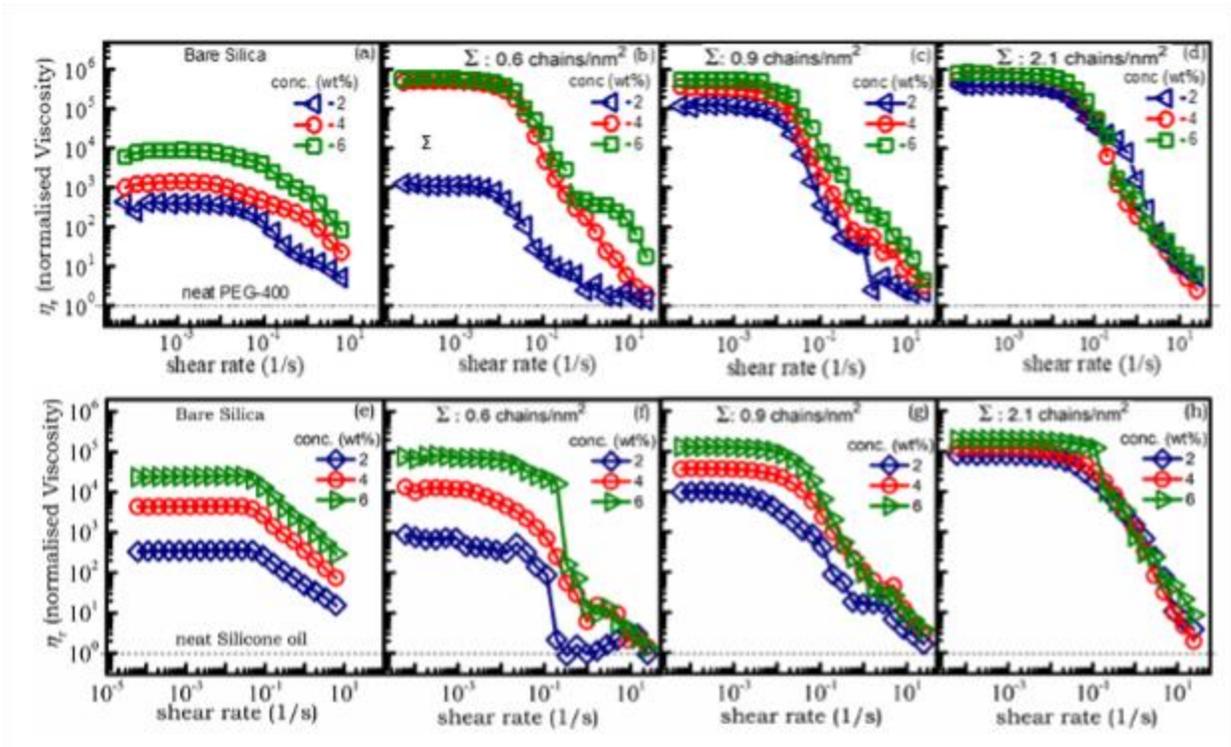

*Figure 2: Normalized viscosity of our GNPs dispersions in PEG400 of different concentrations of bare GNPs in (a), GNPs having grafting density of 0.6 chain/nm$^2$ in (b), 0.9 chain/nm$^2$ in (c), and 2.1 chains/nm$^2$ in (d). Similarly, the normalized viscosity of these GNPs dispersions in low viscosity neat silicone oil for unmodified GNPs in (e) and grafted GNPs having grafting density of 0.6, 0.9, and 2.1 chain/nm$^2$ in (f), (g) and (h) respectively.*

Next, we measure the viscosity of pure liquids – silicone oil and PEG 400 for varying shear rates. Subsequently, we conducted rheological experiments with GNP-modified silicone oil and PEG 400 for different weight fractions of the fillers. The normalized viscosity $\eta_r$, which is the ratio of the composite and pure matrix viscosities as a function of the shear rate is shown in **Figure 2** for both the liquid matrices under the variation of filler loading for different grafting densities. We also measure the liquids' viscosities when loaded with bare NPs. All the experiments are done in ambient conditions. As shown in **Figure 2,** all the samples show shear thinning behavior.

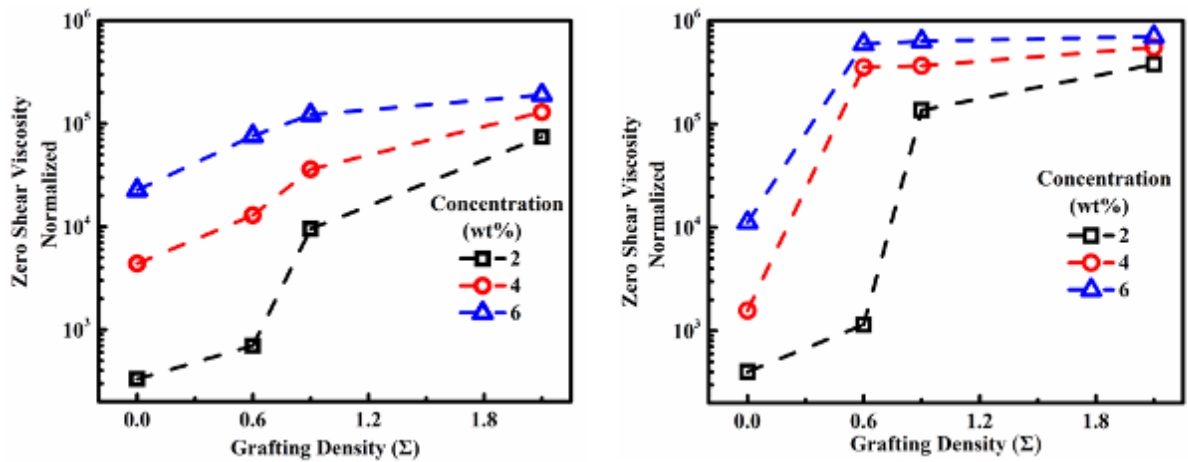

*Figure 3:* The viscosity enhancement is shown as a function of grafting density for different filler concentrations for the silicone oil and PEG 400 matrices in (a) and (b), respectively, for three different loading of GNPs.

The viscosity of all the samples increases with increasing filler loading, which is evident from the upward shifting of flow curves for a given grafting density and base liquid. We select two matrices for the rheological experiments because they exhibit different levels of compatibility with NPs and grafted chains. The chemical moieties of the grafted chains show greater compatibility with PEG 400 than with silicone oil, which allows us to establish a more general mechanism for the phenomenon. In **Figure 2**, we observe a significant increment in the viscosity of GNP loaded liquids – PEG 400 and silicone oil, by 1-3 orders of magnitude compared to bare NPs. Further, the viscosity plateau shifts to higher values with the increment in the grafting density. Finally, for the highest grafting density (~ 2.1 chains/nm$^2$), the viscosity flow curves corresponding to various filler loading almost overlap with each other. Moreover, for the highest grafting density, the viscosity plateau is of the same magnitude for all concentrations for both liquid matrices. The plateau value of normalized zero shear viscosity of all the samples is recapitulated in **Figure 3**. It clearly suggests a wide variability in rheological properties of the composites. We infer that the wider variation in the viscosity enhancement is an attribute of the the ratio of EO and PO units present in our random copolymer, and their complex interaction with the surrounding organic/inorganic matrix. We also note that the effects of grafting density are more pronounced for PEG 400 systems (**Figure 2e-h**), which is evident from the observation that at the moderate grafting density (~ 0.9 chains/nm$^2$), there is an abrupt viscosity jump by 3 orders of magnitude for all the concentrations. However, this jump in viscosity is more gradual for loading of various concentrations when silicone oil is used as a matrix.

To establish the underpinning mechanisms of viscosity enhancement/thickening of liquids, we perform CGMD simulations using a phenomenological model system. In this model system, the free matrix is represented by the finitely extensible nonlinear elastic (FENE) chains[29] of length N=30. The interaction between a pair of monomer is represented by the Lennard-Jones (LJ) potential. The LJ interaction is truncated at $2.5\sigma$ ($\sigma$ is particle diameter) for the non bonded pairs, while the cut-off distance for the bonded monomers is $1.122\sigma$. Further, a NP is modeled as an LJ particle of size $3\sigma$. The grafted chains are also modelled as FENE chains. Each grafted chain consists of 10 monomers. One of the end monomers of a chain is attached to the NP surface.

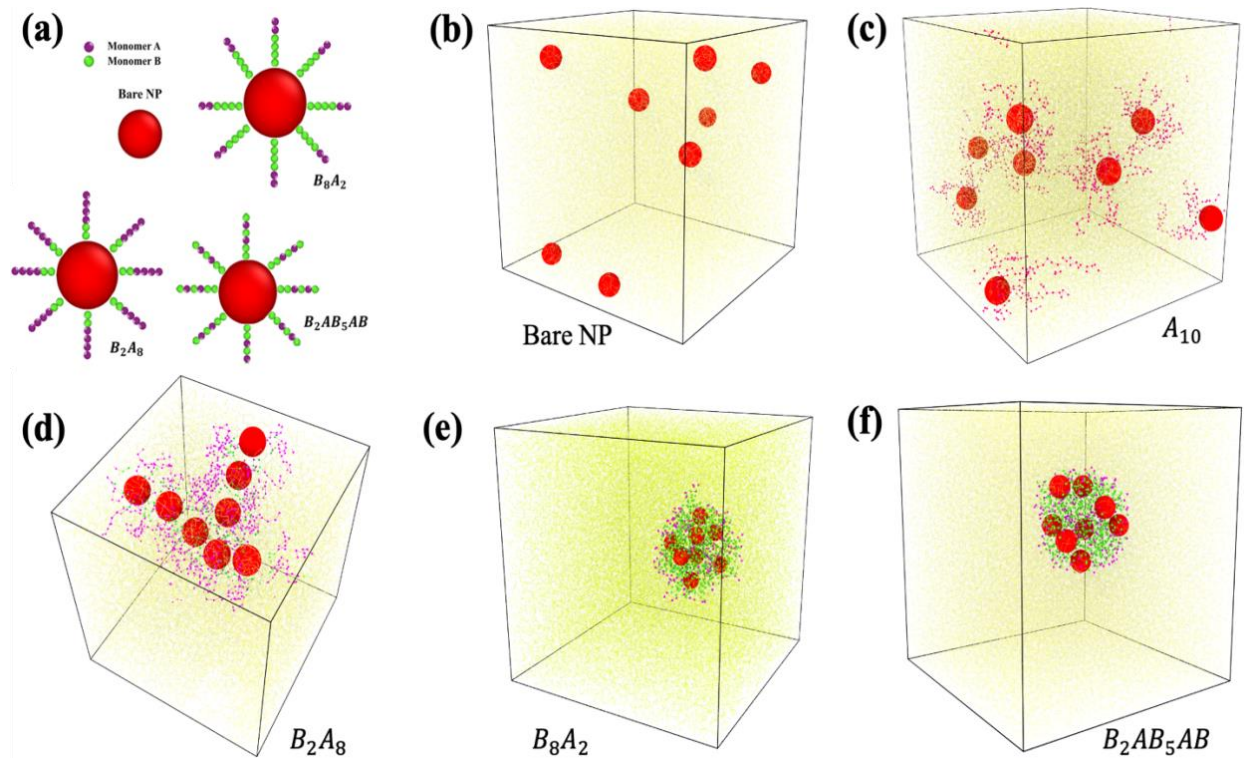

**Figure 4:** CGMD simulations of GNP composites for graft density $\Sigma=0.9\sigma^{-2}$ and NP loading ($\phi=0.02$). (a) Schematic representation of GNP models. The MD snapshots of the composite for bare NPs, $A_{10}$ grafting, $B_2A_8$ grafting, $B_8A_2$ grafting and $B_2AB_5AB$ graftings are shown in in (b), (c), (d), (e) and (f), respectively. The red particles are nanoparticles. The background yellow color represents the free polymer chains matrix. The purple and green correspond to A and B type moieties, respectively.

Schematic representations of the bare NP and GNPs are shown in **Figure 4(a)**. The purple beads (A-type moieties) are compatible with the free matrix and green beads (B-type moieties) are incompatible with the free matrix. The LJ potential also represents the interaction between free chain and grafted chain monomers. The A-type moiety and free chain monomer interact through the LJ interaction with a cut-off distance of 2.5σ. In contrast, the LJ interaction between a pair of B-type moiety and a free chain monomer is truncated at a distance of 1.122σ in order to represent repulsive interaction between them.

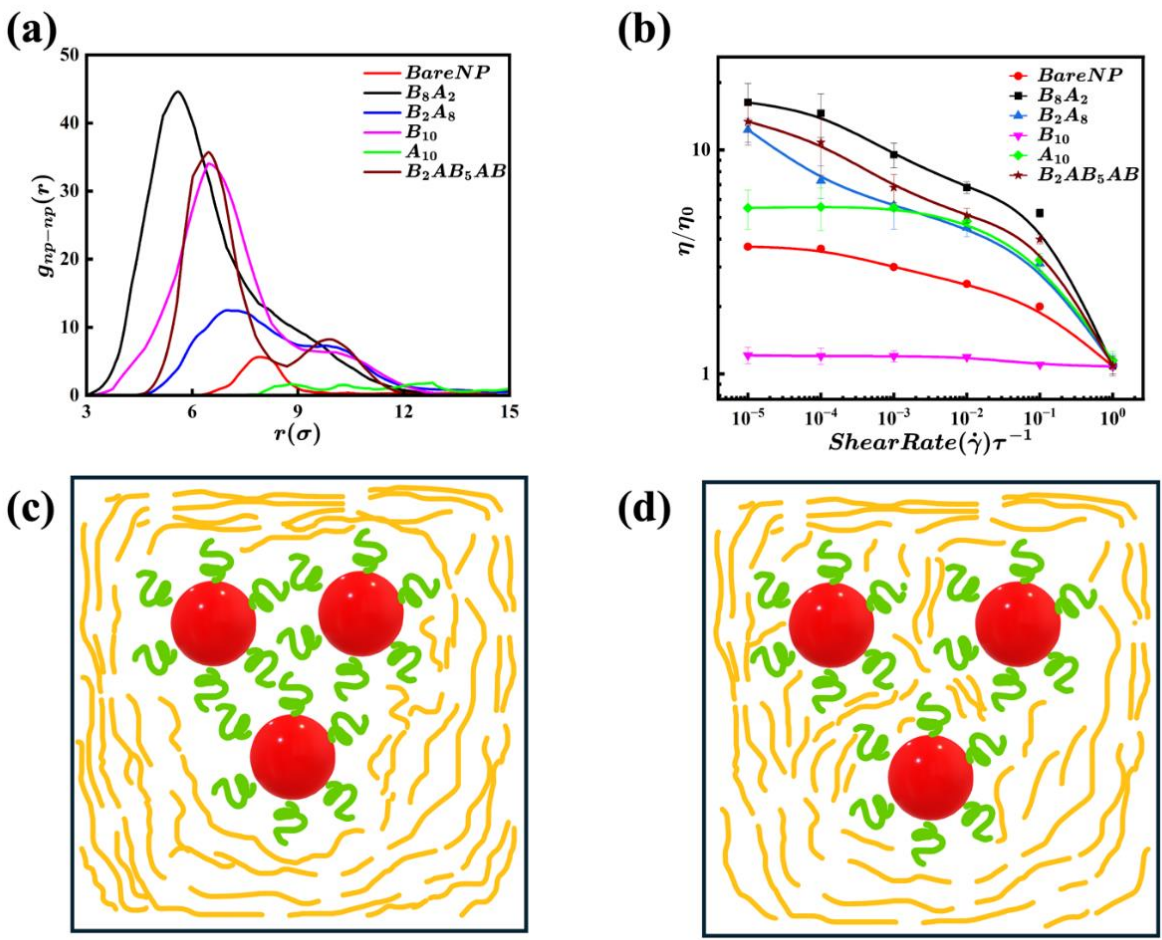

**Figure 5:** NP-NP pair correlation functions, g(r), are shown in (a), and viscosity is plotted as a function of shear rate in (b) for all the simulated systems. The schematic representation of a contact GNP aggregate and a bridged GNP aggregates are shown in (c) and (d), respectively.

The simulations are performed in an isothermal-isobaric ensemble (NPT) within the LAMMPS[30] simulation environment. More details of the model and simulation can be seen in the **SI.** In our simulations, we consider several sequences of copolymers viz., *$A_{10}$, $B_2A_8$, $B_8A_2$ and $B_2AB_5AB$* (random sequence). The first moiety in these sequences is grafted to the NP surface; and the last moiety represents the free end of a graft. MD snapshots of bare NPs and GNPs in the polymer matrix are shown in **Figure 4(b)-(f)** for a grafing density Σ=0.9 and filler loading $\phi = 0.02$. The grafting density $\Sigma$ is defined as the number of polymer chains per unit surface area of a NP. The filler loading ($\varphi$) is calculated as the ratio of GNPs volume to the total volume of the system. These simulations show a wide range of GNP distribution in the polymer matrix depending on the composition and sequence of A and B moieties in graft chains. The corresponding NP-NP pair correlation functions, $g(r)$, and shear dependent viscosity, $\eta$, are shown in **Figure 5**. The viscosities of the composties are normalized by the pure matix visicoy, $\eta_0$. The bare NPs appear to disperse well in the polymer matrix (**Figure 4b**). The extent of dispersion further increases noticeably as the NPs are grafted with pure homopolymers of type-A (**Figure 4c**). This is also evident from **Figure 5a** wherein the the first peak height of $g(r)$ decreases for $A_{10}$ with respect to bare NPs. Also position of this peak is shifted to the right indicating longer interparticle spacing for $A_{10}$. As a consequence, the viscosity increases. In this case, chemicually similar grafting helps improve the dispersion of NPs further and higher viscosity. As we include B moieties in the graft chains, the viscosity of the composites increases further, but NPs tend to aggregate. For the $B_2A_8$, a one-dimensional string-like aggregate forms in the system (**Figure 4d**), and the viscosity is higher than the pure $A_{10}$ case. In this aggregate, the free chains intercalate between NPs. A schematic representation of this scenario is illustrated in **Figure 5d**. Next, we choose a sequence with predominant B moieties. The MD snapshots of the system with $B_8A_2$ grafted NPs are shown in **Figure 4e**, wherein GNPs are microphase separated from the free polymer matrix. The free chains do not intercalate the aggregate. We schematically illustrate this scenario in **Figure 5c**. However, the free end of the grafts contains two A-type monomers that strongly interact with the free polymers around the aggregate. Hence, in spite of microphase separation, the system viscosity is found to be the highest. We now disucss the random sequence case as done in our experiments. For the random sequence, a 3D aggregate of NPs is formed. The viscosity is slightly higher than that of the $B_2A_8$ case, and lower than that of the $B_8A_2$ case. The first peak height of $g(r)$ appears in between the two diblock cases ($B_8A_2$ and $B_2A_8$). The peak position is also in between the two diblock cases. We also simulate a case wherein NPs are grafted with completely dissimilar chains ($B_{10}$). As expected, this leads to a microphase separation (contact aggregate). There is no significant interaction at the liquid - GNP aggregate interface, and the viscosity of the base liquid does not change significantly. Overall, copolymer grafting leads to NP aggregations. As schematically shown in **Figure 5c and d**, there are two types of aggregates are possible. Free chains cannot penetrate contact aggregates. While the bridging of GNP via free polymers leads to bridged aggregates. Such bridging of NPs via polymer layers leads to the enhancement of the potential of mean force between a pair of NPs,[31–33] which is proportional to the $g(r)$ peak height. Hence, the copolymer grafting increases the long range interaction between NP-NP and aid in forming bridged aggregates of NPs within the polymer matrix, akin to previous works on bare nanorods in polymer matrices.[34,35] Hence, by combining compatible and incompatible chemical moieties in graft polymer chains, we are able to tune the microstructure and rheology of composites.

In summary, we have decorated the surface of silica NPs (~10nm) using silylated polyetheramine with chemically dissimilar blocks. The grafting density in our experiments varies from 0.57 to 2.04 chains/nm². By comparing the ϕ value reported in the literature,[27] a steady transition from "mushroom to brush regime" to a "true polymer brush regime" for the functionalized silica particles is noted with an increment in the concentration of the reactive polymer. These copolymer GNPs are found to act as an effective rheological modifier for silicone oil and PEG. When loaded with GNPs, the viscosity of the liquids increases by 1–3 orders of magnitude compared to when they contain bare NPs. Our data suggests identical enhancement of a liquid viscosity for very high grafting density (≥ 2.04 chains/nm²) independent of GNP loading. We have also complemented our rheological experiments with computer simulations, revealing that the compatibility between grafted chains and the surrounding matrix is critical. Simulations show that copolymer grafted NPs form microscopic aggregates. GNP aggregates in the liquid strongly interact with the surrounding matrix due to the presence of chemically similar moieties in the grafts. The sequence of A-type and B-type moieties in grafts, the free ends of the grafts, and graft chain lengths create a complex heterogeneous interface. All of these factors lead to an overall viscosity enhancement of the composites. We expect a wide range of interface compositions depending on these factors. The interface composition determines the interaction between a GNP aggregate and the surrounding liquid medium. Our experiments are conducted for two canonical liquids – silicone oil and PEG. However, we expect similar tunability for other liquids in the presence of copolymer grafted NPs. The study highlights the importance of graft design in achieving versatile GNPs aggregates that can function effectively in contrasting environments. With their ability to enhance stability, these GNPs hold promise for use in lubricants, coatings, and other applications where flow control is critical. The work highlights the potential of carefully designed grafting and chemical compositions of GNPs for controlling the flow behaviors of liquids. We expect such sequence design grafting to be a universal rheological modifier across diverse matrices.

## ASSOCIATED CONTENT

## AUTHOR INFORMATION

**Corresponding Authors**

**Titash Mondal**- Rubber Technology Center, IIT Kharagpur, Kharagpur 721302, India; https://orcid.org/0000-0003-0825-6337; Email: titash@rtc.iitkgp.ac.in

**Manish Kaushal-** Department of Chemical Engineering, IIT Kharagpur, Kharagpur 721302, India; Email: mkaushal@che.iitkgp.ac.in

**Tarak K Patra-** Department of Chemical Engineering, IIT Madras, Madras 600036, India; Email: tpatra@iitm.ac.in

**Authors**

**Prama Adhya-** Department of Chemical Engineering, IIT Kharagpur, Kharagpur 721302, India

**Sachin M. B. Gautham -** Department of Chemical Engineering, IIT Madras, Madras 600036, India

## AUTHOR CONTRIBUTION

**PA:** Experimentation, data interpretation, and writing the original draft (Experimental Part)

**SMBG:** Simulation, data interpretation, and writing the original draft (Simulation Part)

Conceiving the idea**,** supervising, acquiring funds, reviewing, writing, and editing the original draft: Simulation (**TKP**), Rheology (**MK**), and Materials Synthesis (**TM**).

## DATA AVAILABILITY STATEMENT

The data for all the work is available from the corresponding authors based on reasonable requests.

## Notes

The authors declare no competing financial interest.

**Acknowledgment**

The work is made possible by financial support from the SERB, DST, and Gov. of India through a core research grant (CRG/2022/006926), core research grant (CRG/2022/006409), IIT Kharagpur (ISIRD grant (IIT/SRIC/CH/EFF/2018-19/143) and the National Supercomputing Mission's research grant (DST/NSM/R&D_HPC_Applications/2021/40). This research used resources of the Center for Nanoscience Materials, Argonne National Laboratory, which is a DOE Office of Science User Facility supported under contract no. DE-AC02-06CH11357. TKP thanks Abhijit Despande and Madivala G. Basavaraj for the helpful discussion.